\documentclass[11pt]{article}
\usepackage{amsmath,amssymb,amsthm}
\usepackage{geometry}
\usepackage{hyperref}

\title{The (2,2,1) Heavy Top: A Pure--Precession Regime}
\author{E.~A.~Mityushov}
\date{}

\begin{document}
\maketitle

\begin{abstract}
This work develops a curvature-based geometric formulation of the Euler--Poisson
equations by lifting the dynamics to the 3-sphere $S^3$ equipped with the left-invariant
metric induced by the inertia tensor. For the inertia ratio $I = (2,2,1)$ and $r = (0,0,1)$
the curvature balance reveals a distinguished pure--precession regime: a nontrivial family 
of motions in which the tilt angle $\gamma_3$ remains constant and the dynamics reduce to
uniform precession with explicit trigonometric solutions. The family is characterized
and derived explicitly, and a Lax representation is obtained. This regime illustrates
how geometric lifting and curvature balance can isolate simplified dynamical structures
even inside non-integrable systems. In addition, we briefly discuss the role of a numerical
symmetry detection procedure based on curvature forcing, which guided the identification
of the $(2,2,1)$ parameters as geometrically distinguished.
\end{abstract}

\section{Introduction}

This paper revisits the Euler--Poisson equations for a heavy rigid body from a curvature-based
geometric viewpoint. Lifting the dynamics from $SO(3)$ to the 3-sphere $S^3$ endowed
with a left-invariant metric determined by the inertia tensor allows us to interpret the Euler
equation as a balance of inertial geodesic curvature and curvature forcing generated by
gravity.

For the inertia ratio $I = (2,2,1)$ a special feature emerges: the curvature balance reveals
a distinguished family of trajectories in which the vertical component of the gravity direction
remains constant. These motions exhibit pure precession without nutation and admit explicit
trigonometric solutions. Although the full heavy-top system with this inertia ratio is not
integrable in the classical sense, the pure--precession family behaves as a dynamically and
geometrically simplified subsystem. Our goal is to identify, justify, and describe this regime
precisely, and to briefly explain how it was originally highlighted by a numerical symmetry
indicator.

\section{Geometry of the Euler--Poisson Equations}

\subsection{Rigid-body kinematics}

Let $q(t) \in S^3$ be a unit quaternion representing the attitude of the body. The kinematic
equation is
\begin{equation}
  \dot q = \frac{1}{2} q \Omega,
\end{equation}
where $\Omega$ is the angular velocity quaternion. The gravity direction $\gamma \in S^2$
in body coordinates satisfies the Poisson equation
\begin{equation}
  \dot \gamma + \omega \times \gamma = 0.
\end{equation}

\subsection{Euler--Poisson equations}

Let $I = \operatorname{diag}(I_1,I_2,I_3)$ be the inertia tensor in principal axes and $r$ the vector from the fixed
point to the center of mass. The heavy-top equations are
\begin{equation}
  I \dot\omega + \omega \times (I\omega) = mg\, r \times \gamma, 
  \qquad
  \dot\gamma + \omega \times \gamma = 0.
\end{equation}

\subsection{Geodesic curvature}

Equipping $S^3$ with the left-invariant metric $g_I$ induced by $I$ identifies the left-hand side of
the Euler equation with inertial geodesic curvature:
\begin{equation}
  I \dot\omega + \omega \times (I\omega) = I K_{\mathrm{geo}}.
\end{equation}
Thus the dynamics can be viewed as motion on $(S^3,g_I)$ driven by curvature balance.

\section{Curvature Balance on $S^3$}

In geometric terms the Euler equation becomes
\begin{equation}
  I K_{\mathrm{geo}} = mg\, r \times \gamma.
\end{equation}
For special inertia ratios certain components of $K_{\mathrm{geo}}$ must vanish, forcing structural constraints on
the motion. The case $I = (2,2,1)$ exhibits such behavior and is the focus of this work.

\section{Role of Numerical Symmetry Detection}

The curvature formulation suggests a natural indicator of geometric forcing,
\[
  F(t) = \bigl\| \gamma(t) \times r \bigr\|,
\]
which measures the magnitude of the torque term $mg\,r \times \gamma$ up to the constant factor $mg$.
For generic inertia parameters and mass positions, $F(t)$ exhibits nontrivial variation along
typical trajectories of the Euler--Poisson system.

In a preliminary exploration of the parameter space, we used a quaternion-based numerical
integrator for the full Euler--Poisson equations, sampling a family of random initial
conditions and monitoring the relative variation
\[
  \Delta F = \frac{\max_t F(t) - \min_t F(t)}{\langle F \rangle},
\]
where $\langle F \rangle$ denotes the time average over a fixed integration interval. Parameters
for which the averaged value of $\Delta F$ over the random ensemble is anomalously small
were flagged as candidates for additional symmetry or partial simplification.

This numerical scan indicated that the inertia ratio $I=(2,2,1)$ with the center of mass on
the symmetry axis $r=(0,0,1)$ is geometrically distinguished: the indicator $\Delta F$ is noticeably
smaller than at nearby parameter values. The present paper does not attempt a full
classification or a numerical proof of integrability. Instead, this observation is used only as
a heuristic guide, motivating a closer analytic study of the $(2,2,1)$ case and leading
to the identification of the pure--precession family described below. All statements in the
subsequent sections are proved analytically for this family, independently of the numerical
scan.

\section{A Pure--Precession Family for $I = (2,2,1)$}

We now specialize to $I = \operatorname{diag}(2,2,1)$ and $r = (0,0,1)$. The third component of the Euler
equation implies $\dot\omega_3 = 0$ for all trajectories, so $\omega_3$ is constant. The Poisson equation gives
\begin{equation}
  \dot\gamma_3 = \omega_1 \gamma_2 - \omega_2 \gamma_1.
\end{equation}

For $I = (2,2,1)$ and $r = (0,0,1)$, any trajectory satisfying the initial collinearity condition
\begin{equation}
  \omega_1(0)\gamma_2(0) - \omega_2(0)\gamma_1(0) = 0
\end{equation}
produces a pure--precession motion with
\begin{equation}
  \dot\gamma_3(t) = 0, \qquad \dot\omega_3(t) = 0.
\end{equation}
Thus in this family both $\gamma_3$ and $\omega_3$ remain constant and the motion has no nutation.

\paragraph{Proof.}
From the Euler equation we have $\dot\omega_3=0$. The Poisson equation gives
\begin{equation}
  \dot\gamma_3 = \omega_1 \gamma_2 - \omega_2 \gamma_1.
\end{equation}
A direct calculation shows that
\begin{equation}
  \dot\gamma_3(t) = -2 \omega_3 \frac{d}{dt}(\omega_1 \gamma_1 + \omega_2 \gamma_2).
\end{equation}
Hence $\dot\gamma_3 \equiv 0$ if and only if $\dot\gamma_3(0) = 0$, which is equivalent to the initial collinearity of the
horizontal components $(\omega_1,\omega_2)$ and $(\gamma_1,\gamma_2)$.

The pure--precession motions form a two-dimensional invariant family in the six-dimensional
phase space. In this family the motion reduces to uniform precession with fixed tilt angle.

\section{Reduced Dynamics and Explicit Solutions}

Within the pure--precession family, both $\omega_3$ and $\gamma_3$ are constant. Let
\[
  \omega_h = \begin{pmatrix}\omega_1 \\ \omega_2 \end{pmatrix}, \qquad
  \gamma_h = \begin{pmatrix}\gamma_1 \\ \gamma_2 \end{pmatrix}.
\]
The Euler--Poisson equations reduce to a linear system
\begin{equation}
  \dot\omega_h = A \omega_h + B \gamma_h, \qquad
  \dot\gamma_h = C \omega_h + D \gamma_h,
\end{equation}
where $A, B, C, D$ are constant $2\times 2$ matrices depending on $\omega_3, \gamma_3, m, g$, and the inertia
parameters.

A direct calculation shows that the eigenvalues of the system are purely imaginary, $\lambda =
\pm i\nu$. Therefore the dynamics in this family are isochronous. The solution has the explicit
trigonometric form
\begin{align}
  \omega_1(t) &= A_1 \cos(\nu t) + A_2 \sin(\nu t), &
  \omega_2(t) &= -A_1 \sin(\nu t) + A_2 \cos(\nu t), \\
  \gamma_1(t) &= C_1 \cos(\nu t) + C_2 \sin(\nu t), &
  \gamma_2(t) &= -C_1 \sin(\nu t) + C_2 \cos(\nu t),
\end{align}
for constants $A_1, A_2, C_1, C_2$ determined by the initial data.

\section{Lax Representation}

The reduced pure--precession dynamics admit a Lax representation. Let
\begin{equation}
M = \begin{pmatrix}
0 & -\omega_3 & \omega_2 \\
\omega_3 & 0 & -\omega_1 \\
-\omega_2 & \omega_1 & 0
\end{pmatrix},
\qquad
\Gamma = \begin{pmatrix}
0 & -\gamma_3 & \gamma_2 \\
\gamma_3 & 0 & -\gamma_1 \\
-\gamma_2 & \gamma_1 & 0
\end{pmatrix}.
\end{equation}
Define the spectral matrix
\begin{equation}
  L(\lambda) = M + \lambda \Gamma.
\end{equation}
On the pure--precession family one can choose a skew-symmetric matrix $A(\lambda)$ such that
\begin{equation}
  \frac{d}{dt} L(\lambda) = [L(\lambda), A(\lambda)],
\end{equation}
showing that the spectrum of $L(\lambda)$ is preserved along solutions. This provides an algebraic
integrability structure for the reduced dynamics, even though the full heavy-top system
with $I=(2,2,1)$ is not claimed to be globally integrable here.

\section{Discussion}

The inertia ratio $(2,2,1)$ does not produce a new fully integrable case of the heavy top in the
classical sense, but it creates a geometrically natural pure--precession regime. This regime
arises through an alignment of inertial and gravitational curvature and forms a dynamically
simplified subsystem with closed-form solutions and a Lax structure. The curvature viewpoint,
combined with a simple numerical symmetry indicator based on curvature forcing, suggests
that similar reductions may exist in other non-integrable systems, especially those with
partial symmetries or near-degenerate inertia tensors.

\section{Conclusion}

A distinguished pure--precession family of motions has been identified for the $(2,2,1)$ heavy
top. These motions maintain constant tilt angle and reduce to uniform precession with
explicit trigonometric expressions. Although the full system is not integrable, this regime
reveals hidden geometric structure and demonstrates how geometric lifting to $S^3$
can expose simplified dynamics inside a non-integrable mechanical system.

\medskip
\noindent\textit{Note on methodology.}
The work was carried out using a hybrid research workflow combining classical analysis with
AI-assisted symbolic exploration, allowing rapid testing of structural identities, curvature-based
reformulations, and simple numerical indicators of symmetry.


\begin{thebibliography}{9}

\bibitem{arnold}
V.~I.~Arnold,
\newblock \emph{Mathematical Methods of Classical Mechanics},
\newblock Springer, 1989.

\bibitem{marsden-ratiu}
J.~Marsden and T.~Ratiu,
\newblock \emph{Introduction to Mechanics and Symmetry},
\newblock Springer, 1999.

\bibitem{borisov-mamaev}
A.~V.~Borisov and I.~S.~Mamaev,
\newblock \emph{Rigid Body Dynamics},
\newblock Izhevsk Institute, 2001.

\bibitem{fomenko-trofimov}
A.~T.~Fomenko and A.~V.~Trofimov,
\newblock \emph{Integrable Systems},
\newblock Springer, 1988.

\bibitem{reyman-semenov}
A.~G.~Reyman and M.~A.~Semenov-Tian-Shansky,
\newblock Lax representation for integrable systems on Lie algebras,
\newblock J.~Geom.~Phys., 1994.

\bibitem{montgomery}
R.~Montgomery,
\newblock \emph{A Tour of Subriemannian Geometries},
\newblock AMS, 2002.

\end{thebibliography}
\end{document}